\documentstyle[amssymb,epsfig,aps,12pt]{revtex}

\begin{document}
\author{James. W. Dufty}
\address{Department of Physics, University of Florida, Gainesville, FL 32611}
\title{{\bf Shear Stress Correlations in Hard and Soft Sphere Fluids}}
\date{\today }
\maketitle

\begin{abstract}
The shear stress autocorrelation function has been studied recently by
molecular dynamics simulation using the $q^{-n}$ potential for very large $n$. 
The results are analyzed and interpreted here by comparing them to the
shear stress response function for hard spheres. It is shown that the hard
sphere response function has a singular contribution and that this is
reproduced accurately by the simulations for large $n$. A simple model for
the stress autocorrelation function at finite $n$ is proposed, based on the
required hard sphere limiting form.
\end{abstract}

\pacs{}

\section{Introduction}

The qualitative features of simple atomic fluids are often modelled using
the hard sphere pair potential: zero for $q>\sigma $ and infinite for $%
q<\sigma $. It is well-known that the thermodynamic and structural features
of the hard sphere fluid are represented continuously by the soft sphere $%
q^{-n}$ potential in the limit $n\rightarrow \infty $. In contrast, the
dynamical properties for hard and soft sphere fluids are different \ since
the collision time (time required for finite momentum transfer) is strictly
zero in the former case and of order $1/n$ in the latter case. As shown
below, this has a significant effect on the short time behavior of time
correlation functions. Recently, Powles and co-workers \cite%
{heyes,powles1,powles2} have investigated the time correlation functions
determining the shear and bulk viscosity for the $q^{-n}$ fluid at large
values of $n$ (up to $n=288$). The initial values of these functions
diverges $\propto n$, and the simulation results for the normalized
correlation functions appear to scale with the single characteristic time $%
\tau _{n}\propto 1/n$. These results are puzzling when considered as an
approach to the hard sphere limit. The initial values, i.e. the equal time \
stress tensor fluctuations, are related to elastic constants which should be
finite for the hard sphere fluid. Also, the only time scale in the problem $%
\tau _{n}$ vanishes for large $n$ leaving no finite time dynamics for the
hard spheres fluid. In this picture, the finite Green-Kubo shear viscosity
results from the product of the diverging initial condition and a vanishing
correlation time. Clearly, the relationship of properties at large $n$ to
those at $n=\infty $ must be more complex. The objective here is to resolve
these paradoxes by calculating the corresponding properties for the hard
sphere\ fluid directly to allow comparison with those at large $n$.
Attention is restricted to the response function for shear viscosity, but
similar results apply for the bulk viscosity response function as well.

Standard methods of linear response lead to Green-Kubo expressions for the
shear viscosity as the time integral of a shear response function. For soft
spheres this response function is proportional to the autocorrelation
function for the microscopic stress tensor. The form of the stress tensor is
identified from the micrscopic local conservation law for the momentum
density. Since it explicitly involves the interparticle force, a divergence
for large $n$ can be anticipated. Application of the same linear response
methods for the hard sphere fluid lead to several differences:

\begin{itemize}
\item The response function is not simply the stress tensor autocorrelation
function, as for soft spheres, but has an additional singular contribution
proportional to $\delta (t)$.

\item The coefficient of the singular contribution can be calculated exactly
and is simply related to the limiting form of the stress fluctuations for
the soft sphere fluid.

\item The form of the microscopic stress tensor for hard spheres is quite
different from that for soft spheres, and the initial value of the stress
tensor autocorrelation function is finite.

\item The hard sphere stress tensor autocorrelation function has a
characteristic time scale of the mean free time, which is the same as that
for the soft sphere fluid at large $n$.
\end{itemize}

\noindent The singular part of the hard sphere correlation function evolves
continuously from the soft sphere correlation function as $n\rightarrow
\infty $, and is closely related to the divergent initial value. The results
of Powles et al. are shown to be an accurate measurement of this singular
part. However, the resolution of this component in the simulation tends to
mask the finite remainder that persists on the \ mean free path time scale,
whose existence is clear from the hard sphere calculations and which is
essential for the correct density dependence of transport properties. An
alternative representation of the simulation results is proposed to more
closely demonstrate the approach to the hard sphere limit.

The relevant features of the dynamics and statistical mechanics of hard
spheres are introduced in the next section to provide the necessary
definitions and to contrast with the corresponding features for soft
spheres. In Section III the results for linear response to a transverse
velocity field perturbation are quoted, and the Green-Kubo expression is
identified. The singular contribution is evaluated exactly and given a
representation appropriate for comparison with the MD simulation results.
The remainder, determined from the stress tensor autocorrelation function
for hard spheres is evaluated in an Enskog approximation. The limiting form
for soft spheres approaching the hard sphere limit is discussed on the basis
of these results. Some final comments are offered in Section IV.

It is a pleasure to dedicate this brief contribution to Keith Gubbins -
friend, mentor, and colleague - who has played such an important role in
unfolding the mysteries of transport in simple and complex fluids.

\section{Statistical Mechanics of Hard Spheres}

\label{sec2}Any interatomic potential with discontinuities (e.g., hard
sphere, square well) implies an infinite force at the point of
discontinuity. Consequently, the usual formulation of Newtonian dynamics in
terms of forces alone is no longer appropriate. In this section the dynamics
and statistical mechanics for a system of $N$ smooth hard spheres with
diameter $\sigma $ is reviewed briefly \cite{ernst}.

The dynamics consists of free streaming until a given pair $i,j$ is in
contact, at which \ point the relative velocity of that pair changes
instantaneously according to the elastic collision rule 
\begin{equation}
\widetilde{{\bf g}}_{ij}={\bf g}_{ij}-2\widehat{\text{\boldmath$\sigma $}}(%
{\bf g}_{ij}\cdot \widehat{\text{\boldmath$\sigma $}})  \label{2.1}
\end{equation}%
Here ${\bf g}_{ij}={\bf v}_{i}-{\bf v}_{j}$ is the relative velocity for
particles $i$ and $j$. The state of the system at time $t$ is completely
characterized by the positions and velocities of all spheres at that time
and is represented by a point in the associated $6N$ dimensional phase
space, $\Gamma _{t}\equiv \left\{ {\bf q}_{1}(t),..,{\bf q}_{N}(t),{\bf v}%
_{1}(t),..,{\bf v}_{N}(t)\right\} $. The sequence of free streaming and
binary collisions determines uniquely these positions and velocities of the
spheres at time $t$ for given initial conditions. A more complete notation
expressing this dependence on initial conditions is $\Gamma _{t}(\Gamma
_{t^{\prime }})$. Observables of interest are represented by the phase
functions $A(\Gamma _{t}(\Gamma ))$, and their average for given statistical
initial data at $t=0$ is defined by

\begin{equation}
\langle A(t);0\rangle \equiv \int d\Gamma \rho (\Gamma )A(\Gamma _{t}(\Gamma
)),  \label{2.2}
\end{equation}%
where $\rho (\Gamma )$ is the probability density or ensemble for the
initial state, normalized to unity. An equivalent representation of this
average is obtained by changing variables to integrate over $\Gamma _{t}$ \
rather than over $\Gamma $ expressed as a function of $\Gamma _{t}$, i.e.  $%
\Gamma =\Gamma _{t}^{-1}\left( \Gamma _{t}\right) $.  This change of
variables in (\ref{2.2}) allows the time dependence to be expressed in terms
of the probability density 
\begin{equation}
\langle A(t);0\rangle \equiv \int d\Gamma \rho (\Gamma ,t)A(\Gamma )=\langle
A;t\rangle ,  \label{2.3}
\end{equation}%
where the probability density at time $t$ is defined by $\rho (\Gamma
,t)\equiv \rho (\Gamma _{t}^{-1}\left( \Gamma \right) )$. The Jacobian of
the transformation is unity. All of the above is the same for both hard and
soft spheres.

The equilibrium time correlation functions for observables $A$ and $B$ are
defined by 
\begin{equation}
C_{AB}(t)\equiv <\delta A(t)\delta B>_{e}=\int d\Gamma \rho _{e}(\Gamma
)\delta A\left( t\right) \delta B,\hspace{0.3in}\delta X\equiv X-<X>_{e}.
\label{2.4}
\end{equation}%
where $A\left( t\right) =A(\Gamma _{t}(\Gamma _{0}))$. The Gibbs
distribution given by 
\begin{equation}
\rho _{e}(\Gamma )=W(\left\{ q_{ij}\right\} )\prod_{i}\phi \left(
v_{i}\right) ,\hspace{0.3in}\phi \left( v_{i}\right) =\left( \frac{\beta m}{%
2\pi }\right) ^{3/2}e^{-\beta mv^{2}/2}  \label{2.5}
\end{equation}%
is a stationary state for the hard sphere dynamics. The ''overlap'' function 
$W(\left\{ q_{ij}\right\} )$ is defined to vanish if any pair overlap, $%
q_{ij}<\sigma $, and is normalized to unity when integrated over all
configuration space. The equilibrium correlation function is stationary 
\begin{equation}
<\delta A(t)\delta B>_{e}=<\delta A\delta B(-t)>_{e}  \label{2.6}
\end{equation}%
where $B(-t)=B(\Gamma _{-t}(\Gamma ))$ has the time reversed dynamics.

For practical purposes it is useful to identify the generators $L_{\pm }$
and $\overline{L}$ \ for the two representations, defined for $t>0$ by 
\begin{equation}
A(\pm t)\equiv e^{\pm L_{\pm }t}A(\Gamma ),\hspace{0.3in}\rho (\Gamma
,t)=e^{-\overline{L}t}\rho (\Gamma ).  \label{2.7}
\end{equation}%
These are not the usual generators of Hamilton's equations for continuous
forces and are somewhat more complex due to the singular nature of hard
spheres. They are given by 
\begin{equation}
L_{\pm }=\sum_{i=1}^{N}{\bf v}_{i}\cdot {\bf \nabla }_{i}\pm \frac{1}{2}%
\sum_{i=1}^{N}\sum_{j\neq i}^{N}T_{\pm }(i,j),\hspace{0.3in}\overline{L}%
=\sum_{i=1}^{N}{\bf v}_{i}\cdot {\bf \nabla }_{i}-\frac{1}{2}%
\sum_{i=1}^{N}\sum_{j\neq i}^{N}\overline{T}(i,j).  \label{2.10}
\end{equation}%
The first terms on the right sides generate free streaming while the second
terms describe velocity changes. The three binary collision operators, $%
T_{\pm }(i,j)$ and $\overline{T}(i,j)$, for particles $i$ and $j$ are given
by 
\begin{equation}
T_{\pm }(i,j)=\mp \sigma ^{2}\int \ d\Omega \ \Theta (\mp {\bf g}_{ij}{\bf %
\cdot }\widehat{\text{\boldmath$\sigma $}})({\bf g}_{ij}{\bf \cdot }\widehat{%
\text{\boldmath$\sigma $}})\delta ({\bf q}_{ij}-\text{\boldmath$\sigma $}%
)(b_{ij}-1),  \label{2.11}
\end{equation}%
\begin{equation}
\overline{T}(i,j)=\sigma ^{2}\int d\Omega \ \Theta ({\bf g}_{ij}{\bf \cdot }%
\widehat{\text{\boldmath$\sigma $}})({\bf g}_{ij}{\bf \cdot }\widehat{\text{%
\boldmath$\sigma $}})[\delta ({\bf q}_{ij}-\text{\boldmath$\sigma $}%
)b_{ij}-\delta ({\bf q}_{ij}+\text{\boldmath$\sigma $})],  \label{2.12}
\end{equation}%
where $d\Omega $ denotes the solid angle integration for the unit vector $%
\widehat{\text{\boldmath$\sigma $}}$, ${\bf r}$ is the relative position
vector of the two particles, and the operator $b_{ij}$ is a substitution
operator, $b_{ij}F({\bf g}_{ij}){\bf =}F(b_{ij}{\bf g}_{ij})$, which changes
the relative velocity ${\bf g}_{ij}={\bf v}_{i}-{\bf v}_{j}$ into its
scattered velocity according to (\ref{2.3}) $b_{ij}{\bf g}_{ij}=\widetilde{%
{\bf g}_{ij}}$. For continuous potentials all generators are the same $%
L_{\pm }=\overline{L}$ but for hard spheres all three generators clearly are
different.

The local microscopic conservation laws follow from the above dynamics by
differentiating the local conserved densities of mass, energy, and momentum
with respect to time. The relevant conservation law for the discussion here
is that for momentum density 
\begin{equation}
\partial _{t}p_{\alpha }({\bf r},\pm t)+\partial _{j}\tau _{\pm \alpha \beta
}({\bf r},\pm t)=0,  \label{2.13}
\end{equation}
\begin{equation}
{\bf p}({\bf r})=\sum_{i=1}^{N}{\bf p}_{i}\delta ({\bf r}-{\bf q}_{i}),%
\hspace{0.5cm}\partial _{j}\tau _{\pm \alpha \beta }({\bf r})=\mp L_{\pm
}p_{\alpha }({\bf r})  \label{2.14}
\end{equation}
The subscript $\pm $ distinguishes the forms of the momentum flux for the
case $t>0$ and $t<0$. The detailed expression for $\tau _{\pm ij}({\bf r},t)$
is not required below, only its volume integral 
\begin{eqnarray}
{\cal T}_{\pm \alpha \beta } &\equiv &\int d{\bf r}\tau _{\pm \alpha \beta }(%
{\bf r})  \nonumber \\
&=&\sum_{i=1}^{N}mv_{i,\alpha }v_{j,\beta }+\frac{1}{2}\sigma
^{3}\sum_{i\neq j}^{N}\int d\Omega \ \Theta (\mp {\bf g}_{ij}{\bf \cdot }%
\widehat{\text{{\bf $\sigma $}}})m({\bf g}_{ij}{\bf \cdot }\widehat{\text{%
{\bf $\sigma $}}})^{2}\widehat{\sigma }_{\alpha }\,\widehat{\sigma }_{\beta
}\delta ({\bf q}_{ij}-\text{{\bf $\sigma $}})  \label{2.15}
\end{eqnarray}
This form for the hard sphere momentum flux(es) can be contrasted with that
for the soft sphere potential $v(q)=\epsilon \left( \sigma /q\right) ^{n}$%
\begin{equation}
{\cal T}_{\alpha \beta }=\sum_{i=1}^{N}mv_{i,\alpha }v_{i,\beta }+\frac{1}{2}%
\sum_{i\neq j}^{N}\widehat{q}_{ij\alpha }\,\widehat{q}_{ij\beta }n\epsilon
\left( \frac{\sigma }{q_{ij}}\right) ^{n}  \label{2.16}
\end{equation}
which can be written for large $n$ as 
\begin{equation}
{\cal T}_{\alpha \beta }=\sum_{i=1}^{N}mv_{i,\alpha }v_{i,\beta }+n\frac{1}{2%
}\sigma \epsilon \sum_{i\neq j}^{N}\int d\Omega \ \widehat{\sigma }_{\alpha
}\,\widehat{\sigma }_{\beta }\delta ({\bf q}_{ij}-\text{{\bf $\sigma $}})
\label{2.17}
\end{equation}
While (\ref{2.15}) and (\ref{2.17}) are similar, both involving a surface
integral for pairs of particles at contact, there are two significant
differences. The hard sphere stress tensor is momentum dependent and the
soft sphere stress tensor is proportional to $n$, diverging in the large $n$
limit.

\section{Linear Response}

\label{sec3}The results of the last section now allow a discussion of the
response functions for the hard sphere fluid. Attention here is restricted
to the response to an initial transverse perturbation of the velocity field, 
$u_{y}(x,0)=A\sin (x/\lambda )$. The velocity field for $\lambda $ much
larger than the mean free path and for times long compared to the mean free
time then obeys the diffusion equation 
\begin{equation}
\partial _{t}u_{y}(x,t)+\left( \eta /n_{e}m\right) \nabla ^{2}u_{y}(x,t)=0
\label{3.2}
\end{equation}%
where $\eta $ is the shear viscosity and $n_{e}$ is the equilibrium density
. The derivation of this result is straightforward using the standard
methods of linear response and the Liouville dynamics defined in the last
section. The details will be given elsewhere \cite{dufty} and only \ the
results discussed here. The shear viscosity is given by the Green-Kubo
expression 
\begin{equation}
\eta =\beta \left[ -\frac{1}{V}\left\langle {\cal T}_{+xy}\Lambda
_{xy}\right\rangle _{e}+\int_{0}^{\infty }dt\frac{1}{V}\left\langle {\cal T}%
_{+xy}\left( t\right) {\cal T}_{-xy}\right\rangle _{e}\right]   \label{3.3}
\end{equation}%
where 
\begin{equation}
\Lambda _{xy}=\int d{\bf r}xp_{y}({\bf r})=\sum_{i=1}q_{ix}mv_{iy}
\label{3.3a}
\end{equation}%
For soft spheres ${\cal T}_{+xy}={\cal T}_{-xy}={\cal T}_{xy}$ $\ $and $%
\left\langle {\cal T}_{xy}\Lambda _{xy}\right\rangle _{e}=0$ since ${\cal T}%
_{xy}$ is independent of the particle velocities while $\Lambda _{xy}$ is an
odd function of the \ velocities. Thus, the usual Green-Kubo form for the
shear viscosity in terms of the time integral of the stress tensor
autocorrelation function is regained 
\begin{equation}
\eta ^{ss}=\beta \int_{0}^{\infty }dt\frac{1}{V}\left\langle {\cal T}%
_{xy}\left( t\right) {\cal T}_{xy}\right\rangle _{e}  \label{3.4}
\end{equation}%
In contrast, $\left\langle {\cal T}_{+xy}\Lambda _{xy}\right\rangle _{e}$
does not vanish for hard spheres. The equivalence of (\ref{3.3}) and  (\ref%
{3.4}) in the limit of $n\rightarrow \infty $ requires 
\begin{equation}
\lim_{n\rightarrow \infty }\;\frac{1}{V}\left\langle {\cal T}_{xy}\left(
t\right) {\cal T}_{xy}\right\rangle _{e}=-\frac{2}{V}\left\langle {\cal T}%
_{+xy}\Lambda _{xy}\right\rangle _{e}\delta (t)+\frac{1}{V}\left\langle 
{\cal T}_{+xy}\left( t\right) {\cal T}_{-xy}\right\rangle _{e}  \label{3.5a}
\end{equation}%
This is the primary observation of this work regarding the relationship of
shear response for soft and hard spheres. The hard and soft sphere stress
autocorrelation functions differ by a singular term.

To explore this relationship, two results are noted. First, for hard spheres
the coefficient of the singular term can be calculated exactly to get \cite%
{dufty} 
\begin{equation}
-\frac{2}{V}\left\langle {\cal T}_{+xy}\Lambda _{xy}\right\rangle _{e}=\frac{%
4}{5}\sigma \left( \frac{m}{\beta \pi }\right) ^{1/2}\left( p-p^{k}\right) .
\label{3.6}
\end{equation}%
Also, for soft spheres the exact initial condition is \cite{zwanzig} 
\begin{equation}
\frac{1}{V}\left\langle {\cal T}_{xy}\left( 0\right) {\cal T}%
_{xy}\right\rangle _{e}\equiv \frac{1}{\beta }G_{\infty }=\frac{1}{\beta }%
\left[ p^{k}+\frac{1}{5}\left( n-3\right) \left( p-p^{k}\right) \right] .
\label{3.7}
\end{equation}%
Here $\beta =1/k_{B}T$, $p^{k}=n_{e}/\beta $, $p$ is the pressure 
\begin{equation}
p=p^{k}\left( 1+\frac{2}{3}\pi n_{e}^{\ast }\chi \right) ,  \label{3.8}
\end{equation}%
and $\chi $ is the equilibrium pair correlation function evaluated at $%
q_{ij}=\sigma $. Equation (\ref{3.7}) shows the divergence of the high
frequency shear modulus, $G_{\infty }$, an $n\rightarrow \infty $.
Furthermore, the dominant short time behavior of $\left\langle {\cal T}%
_{xy}\left( t\right) {\cal T}_{xy}\right\rangle _{e}$ for soft spheres with
large $n$ is 
\begin{equation}
\left\langle {\cal T}_{xy}\left( t\right) {\cal T}_{xy}\right\rangle
_{e}=\left\langle {\cal T}_{xy}\left( 0\right) {\cal T}_{xy}\right\rangle
_{e}\left[ 1-\left( \frac{t}{\tau _{n}}\right) ^{2}+\text{ order }t^{4}%
\right]   \label{3.9}
\end{equation}%
with the characteristic short time scale 
\begin{equation}
\tau _{n}=\frac{\sigma \sqrt{\beta m}}{n}=\frac{\tau _{1}}{n}  \label{3.10}
\end{equation}%
Comparison of (\ref{3.6}) and (\ref{3.7}) shows the relationship 
\begin{equation}
\lim_{n\rightarrow \infty }\frac{4\tau _{n}}{\sqrt{\pi }}\frac{1}{V}%
\left\langle {\cal T}_{xy}\left( 0\right) {\cal T}_{xy}\right\rangle
_{e}=\left( -\frac{2}{V}\left\langle {\cal T}_{+xy}\Lambda
_{xy}\right\rangle _{e}\right)   \label{3.11}
\end{equation}%
Note that the left side is finite in this limit due to the factor of $\tau
_{n}\approx 1/n$.

These results provide an appropriate means to discuss and interpret the
simulation results of reference \cite{powles2}. The stress tensor
autocorrelation function is written as 
\begin{equation}
\left\langle {\cal T}_{xy}\left( t\right) {\cal T}_{xy}\right\rangle
_{e}=\left\langle {\cal T}_{xy}\left( 0\right) {\cal T}_{xy}\right\rangle
_{e}C_{n}(t)  \label{3.12}
\end{equation}%
and the simulation data is reported for the normalized correlation function $%
C_{n}(t)$ (a subscript $n$ has been included since the simulations are
performed for $n=18,36,72,144,$ and $288$). It is found that the results are
globally well-fit by 
\begin{equation}
\left[ C_{n}(t)\right] _{sim}\approx \text{sech}\left( a\frac{t}{\tau _{n}}%
\right) .  \label{3.13}
\end{equation}%
where $a=\pi ^{3/2}/4\approx 1.4$ (in reference \cite{powles2} $a=\sqrt{2}$%
). This is the puzzling result mentioned in the Introduction above. The
coefficient of $C_{n}(t)$ in (\ref{3.12}) diverges for large $n$ while $%
\left[ C(t)\right] _{sim}$ is a ''universal'' function of time with a
vanishing time scale. To understand this paradoxical behavior combine (\ref%
{3.11}) - (\ref{3.13}) to get, for large $n$%
\begin{eqnarray}
\left[ \frac{1}{V}\left\langle {\cal T}_{xy}\left( t\right) {\cal T}%
_{xy}\right\rangle _{e}\right] _{sim} &=&\left( -\frac{2}{V}\left\langle 
{\cal T}_{+xy}\Lambda _{xy}\right\rangle _{e}\right) \frac{\sqrt{\pi }}{%
4\tau _{n}}\text{sech}\left( a\frac{t}{\tau _{n}}\right)   \nonumber \\
&=&\left( -\frac{2}{V}\left\langle {\cal T}_{+xy}\Lambda _{xy}\right\rangle
_{e}\right) \delta _{n}\left( t\right)   \label{3.14}
\end{eqnarray}%
In the second line $\delta _{n}(t)$ is a representation of the delta
function as $n\rightarrow \infty $ defined by 
\begin{equation}
\delta _{n}\left( t\right) =\frac{a}{\pi \tau _{n}}\text{sech}\left( a\frac{t%
}{\tau _{n}}\right)   \label{3.15}
\end{equation}%
It is now clear that the MD simulation has measured the singular
contribution on the right side of (\ref{3.5a}). This reinterpretation of the
MD results shows a remarkable confirmation of the hard sphere singular
contribution evolving continuously from the soft sphere stress
autocorrelation function. It also demonstrates the role of the diverging
shear modulus in constructing the amplitude of $\delta _{n}\left( t\right) $%
. Finally, the paradox of the vanishing time scale is now understood as
well. The second term on the right side of (\ref{3.5a}) is expected to have
a time scale of the order of the mean free time (see below), but has not
been well-resolved by the simulations. The normalization to define $C(t)$,
appropriate to identify this singular term, makes observation of the second
term difficult since it then vanishes as $n^{-1}$. Still, such deviations
from the fit (\ref{3.13}) are observed in the simulations, as noted below.

The second term in (\ref{3.5a}) is the hard sphere stress autocorrelation
function. It can be evaluated approximately from Enskog kinetic theory. The
details will be given elsewhere, and only the result quoted here 
\begin{equation}
\frac{1}{V}\left\langle {\cal T}_{+xy}\left( t\right) {\cal T}%
_{-xy}\right\rangle _{e}\rightarrow \beta ^{-1}p^{k}\left( 1+\frac{2}{5}%
\frac{p-p^{k}}{p_{k}}\right) ^{2}e^{-t/\tau _{mfp}}  \label{3.17}
\end{equation}%
where $\tau _{mfp}$ is the mean free time 
\begin{equation}
\tau _{mfp}=\frac{5}{24}\frac{p_{k}}{\left( p-p^{k}\right) }\sqrt{\pi }\tau
_{1}.  \label{3.18}
\end{equation}%
The pressures for the \ hard and soft sphere fluids are continuously related
for large $n$ so both the non-singular part in (\ref{3.17}) and the mean
free path should be similar in both cases. It is easily verified that use of
(\ref{3.17}) and (\ref{3.6}) in the Green-Kubo expression (\ref{3.3}) yields
the exact Enskog shear viscosity. Both the singular term and the stress auto
correlation function contribute at intermediate velocities. In the low
density limit the singular contribution to the viscosity vanishes and only
the stress autocorrelation function contributes. Also, note that the value
at $t=0$ for the hard sphere stress autocorrelation function is finite, in
contrast to that for the soft sphere correlation function. The implications
of this for elastic constants will be discussed elsewhere.

The hard sphere results suggest how to write an approximate stress
autocorrelation function for the soft sphere potential at large $n$. There
should be two contributions with two different time scales. The first is
that which has been determined from the simulations and which yields the
singular contribution \ in the hard sphere limit. It also gives entirely the
exact short time behavior. The second term should vanish at short times, but
tend toward the form (\ref{3.17}) for $t>>\tau _{n}$. A simple approximation
consistent with the above exact properties and which has the proper hard
sphere limit is 
\begin{equation}
\frac{1}{V}\left\langle {\cal T}_{xy}\left( t\right) {\cal T}%
_{xy}\right\rangle _{e}\approx \frac{1}{\beta }G_{\infty }\text{sech}\left( a%
\frac{t}{\tau _{n}}\right) +X(\frac{t}{\tau _{n}})\frac{p^{k}}{\beta }\left(
1+\frac{2}{5}\frac{p-p^{k}}{p_{k}}\right) ^{2}e^{-t/\tau _{mfp}}.
\label{3.19}
\end{equation}%
Equivalently, the normalized correlation function is%
\begin{equation}
C_{n}\left( t\right) \approx \text{sech}\left( a\frac{t}{\tau _{n}}\right)
+\Delta _{n}\left( t\right)   \label{3.21}
\end{equation}%
where $\Delta _{n}\left( t\right) $ is the contribution from the
non-singular part 
\begin{equation}
\Delta _{n}\left( t\right) =X(\frac{t}{\tau _{n}})\frac{p^{k}}{G_{\infty }}%
\left( 1+\frac{2}{5}\frac{p-p^{k}}{p_{k}}\right) ^{2}e^{-t/\tau _{mfp}}
\label{3.21a}
\end{equation}%
The function $X(t)$ controls the crossover from the time scale $\tau _{n}$
to the mean free time scale. For consistency it must be even in time, vanish
up through order $t^{2}$, and approach unity for large times. One simple
form with these properties is 
\begin{equation}
X(t)=\left( \frac{t^{2}}{1+t^{2}}\right) ^{2}  \label{3.22}
\end{equation}%
\begin{figure}[tbh]
\centerline{\epsfig{file=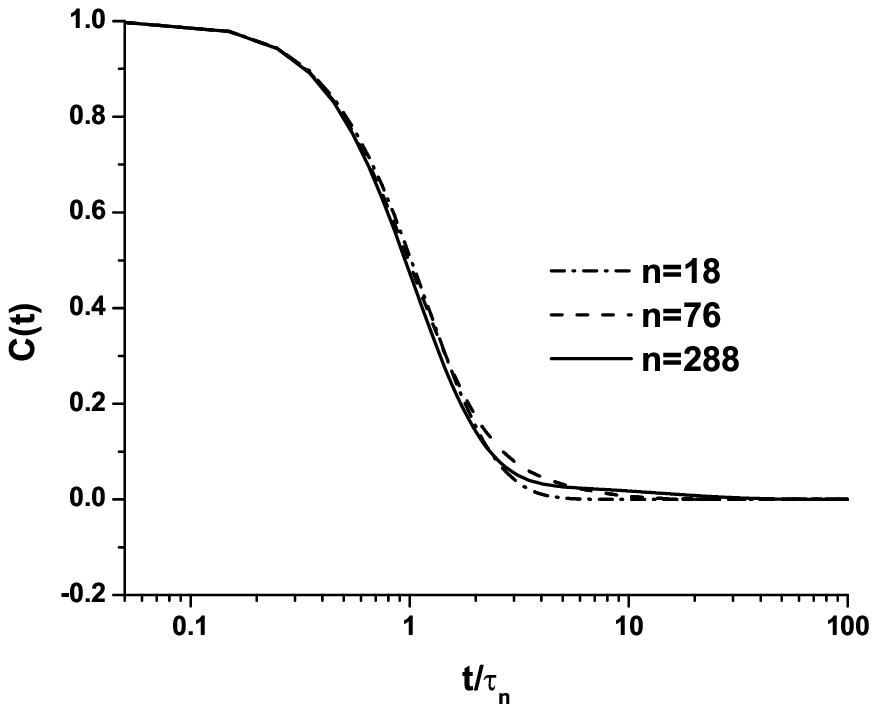,width=0.55\textwidth}}
\caption{Plot of normalized stress autocorrelation function $C(t)$ vs log $t$ at
packing fraction 0.45 for $n = 18$, $76$, and $288$}.
\label{Fig1}
\centerline{\epsfig{file=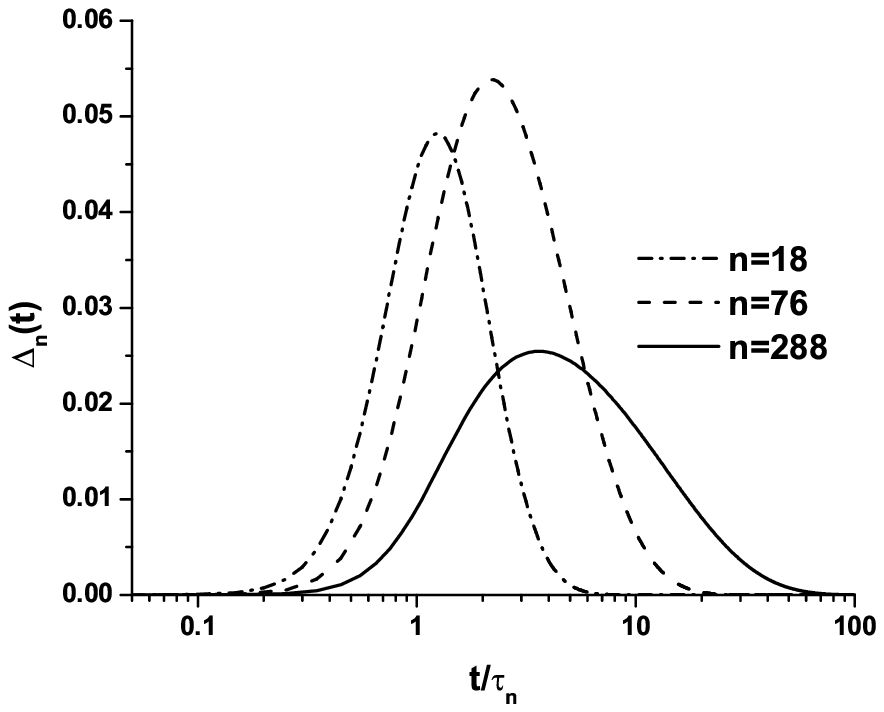,width=0.55\textwidth}}
\caption{Same as Fig1 for $\Delta (t)$, non-singular contribution to $C(t)$}.
\label{Fig2}
\end{figure}
For illustration, Figure 1 shows $C_{n}\left( t\right) $ as a function of $%
t/\tau _{n}$ for $n=18,76,$ and $288$ at the packing fraction $\pi
n_{e}\sigma ^{3}/6=0.45$. This is quite similar to the simulation results in
Figure 5 of reference \cite{powles2}. The small variations with $n$ for $%
t/\tau _{n}>1$ are due to $\Delta _{n}\left( t\right) $. The latter is shown
in Figure 2.

A more rational choice for the crossover function $X(t)$ is obtained by
relating it to the low density autocorrelation function%
\begin{equation}
X(\frac{t}{\tau _{n}})=e^{t/\tau _{mfp}^{(0)}}\left[ C^{(0)}\left( t\right) -%
\text{sech}\left( a\frac{t}{\tau _{n}^{(0)}}\right) \right]   \label{3.22a}
\end{equation}
The superscript $(0)$ denotes the low density limit. For example, $\tau
_{n}^{(0)}$ is the time scale obtained from the expansion of the low density
autocorrelation function to second order in $t$. With this choice the
normalized stress autocorrelation function becomes%
\begin{equation}
C_{n}\left( t\right) \approx \text{sech}\left( a\frac{t}{\tau _{n}}\right) +%
\left[ C^{(0)}\left( t\right) -\text{sech}\left( a\frac{t}{\tau _{n}^{(0)}}%
\right) \right] \frac{p^{k}}{G_{\infty }}\left( 1+\frac{2}{5}\frac{p-p^{k}}{%
p_{k}}\right) ^{2}e^{-t/\tau }.  \label{3.23}
\end{equation}%
where the new time scale of the second term is 
\begin{equation}
\frac{1}{\tau }=\frac{1}{\tau _{mfp}}-\frac{1}{\tau _{mfp}^{(0)}}=\frac{24}{5%
\sqrt{\pi }\tau _{1}}\left( \frac{p-p^{k}}{p_{k}}\right)   \label{3.24}
\end{equation}%
The calculation of $C^{(0)}\left( t\right) $ for all times is still a
difficult problem, but one that has been reduced to two particle dynamics.
The remaining terms are all expressed in terms of the pressure. It should be
interesting to see how well (\ref{3.23}) predicts the density dependence of
the viscosity for the soft sphere fluid.

\section{Discussion}

On purely physical grounds it is clear that the physical properties of the
soft sphere fluid for large $n$ should approach those of the hard sphere
fluid, in general. This requires that the continuous Newtonian dynamics of
the soft sphere fluid must generate the singularities that are inherent in
the dynamics of the hard sphere fluid. The molecular dynamics simulation
results of references \cite{heyes,powles1,powles2} provide the first direct
demonstration of how such singular properties can be extracted from
competition between diverging fluctuations (the high frequency shear
modulus) and a vanishing collision time. More generally, the autocorrelation
functions for the soft sphere fluid have contributions on two time scales, $%
\tau _{n}$ and $\tau _{mfp}$. For asymptotically short times only the former
contributes. There is then a crossover to longer times when the latter
dominates. As the density increases, $\tau _{mfp}$ decreases and if $n$ is
not large the two times can become comparable and the dynamics is a
composition of both. This is the case in Figure 2 for $1\lessapprox t/\tau
_{n}\lessapprox 10$. A reevaluation of the simulation data to extract the
non-singular dynamics on the time scale $\tau _{mfp}$ is of some interest
and could be compared directly to hard sphere simulations at $t>0$. It
appears that a recently proposed approximation \cite{nettleton} to
interpolate correlation functions between their exact short time dynamics
and long time exponential decay does not work in this case due to the
divergent initial value.

Similar questions regarding the relationship of hard and soft sphere systems
can be raised for the solid state. The elastic constants are usually related
to the stress tensor fluctuations which diverge for the soft sphere fluid
but are finite for the hard sphere fluid. To clarify this, the exact
relations (\ref{3.5a}) and (\ref{3.11}) can be combined to give 
\begin{equation}
\lim_{n\rightarrow \infty }\;\left[ \frac{1}{V}\left\langle {\cal T}%
_{xy}\left( t\right) {\cal T}_{xy}\right\rangle _{e}-\frac{1}{V}\left\langle 
{\cal T}_{xy}\left( 0\right) {\cal T}_{xy}\right\rangle _{e}\frac{4\tau _{n}%
}{\sqrt{\pi }}\delta _{n}\left( t\right) =\right] =\frac{1}{V}\left\langle 
{\cal T}_{+xy}\left( t\right) {\cal T}_{-xy}\right\rangle _{e}  \label{4.1}
\end{equation}%
The right side is finite for $t=0^{+}$(\ref{3.23}), which suggests that the
proper elastic properties of the soft sphere fluid require subtraction of
the short time dynamics with the diverging amplitude. This can be viewed as
an annealing process on the time scale $\tau _{n}$.

The analysis of simulation data for thermal conductivity and bulk viscosity
for the soft sphere fluid leads to similar results. The description will be
given in a more detailed report on this topic elsewhere \cite{dufty}.

\section{Acknowledgements}

The author is indebted to J. Powles and D. Heyes for communication of
results prior to publication. This research was supported in part by
National Science Foundation grant PHY 9722133.


\begin{references}
\bibitem{heyes} D. Heyes and J. G. \ Powles , Molec. Phys. {\bf 95}, 259
(1998).

\bibitem{powles1} J. G. \ Powles, G. Rickayzen, and D. Heyes, Proc. R. Soc.
Lond. A {\bf 455}, 3725 (1999).

\bibitem{powles2} J. G. \ Powles and D. Heyes, Mol. Phys. {\bf 98}, 917
(2000).

\bibitem{ernst} M. Ernst, J. Dorfman, W. Hoegy, and J. van Leeuwen, Physica
45, 127 (1969); J. Sengers, M. Ernst, and D. Gillespie, J. Chem. Phys. {\bf %
56}, 5583 (1972).

\bibitem{dufty} J. Dufty, (to be published).

\bibitem{zwanzig} R. Zwanzig and R. Mountain, J. Chem. Phys. 43, 4464 (1965).

\bibitem{nettleton} R. Nettleton, J. Phys. A: Math. Gen.{\bf \ 33,} 7555
(2000).
\end{references}
\end{document}